# Performance Comparison of Minimum Hop vs. Minimum Edge Based Multicast Routing under Different Mobility Models for Mobile Ad hoc Networks

Natarajan Meghanathan
Associate Professor
Department of Computer Science
Jackson State University
Jackson, MS 39217, USA
Phone: 601-979-3661; Fax: 601-979-2478
E-mail: natarajan.meghanathan@jsums.edu

**Abstract**

The high-level contribution of this paper is to establish benchmarks for the minimum hop count per source-receiver path and the minimum number of edges per tree for multicast routing in mobile ad hoc networks (MANETs) under different mobility models. In this pursuit, we explore the tradeoffs between these two routing strategies with respect to hop count, number of edges and lifetime per multicast tree with respect to the Random Waypoint, City Section and Manhattan mobility models. We employ the Breadth First Search algorithm and the Minimum Steiner Tree heuristic for determining a sequence of minimum hop and minimum edge trees respectively. While both the minimum hop and minimum edge trees exist for a relatively longer time under the Manhattan mobility model; the number of edges per tree and the hop count per source-receiver path are relatively low under the Random Waypoint model. For all the three mobility models, the minimum edge trees have a longer lifetime compared to the minimum hop trees and the difference in lifetime increases with increase in network density and/or the multicast group size. Multicast trees determined under the City Section model incur fewer edges and lower hop count compared to the Manhattan mobility model.

## 1 Introduction

A mobile ad hoc network (MANET) is a dynamic distributed system of wireless nodes that move independent of each other in an autonomous fashion. The network bandwidth is limited and the medium is shared. As a result, transmissions are prone to interference and collisions. The battery power of the nodes is constrained and hence nodes operate with a limited transmission range, often leading to multi-hop routes between any pair of nodes in the network. Communication structures (e.g.., paths, trees, connected dominating sets and etc) for routing in wireless ad hoc networks could be principally based on two different approaches [1]: Optimum Routing Approach (ORA) and Least Overhead Routing Approach (LORA). With ORA, the communication structure used at any time instant is always the optimum with respect to a particular metric. On the other hand, with LORA, a communication structure determined for optimality with respect to a particular metric at a time instant is used in the subsequent time instants as long as the communication structure exists. For dynamically changing, distributed, resource-constrained MANETs, the LORA strategy is often preferred over the ORA strategy to avoid the communication overhead incurred in determining the optimum communication structure at every time instant. Hence, we focus on using LORA for the rest of this paper.

Multicasting in ad hoc wireless networks has numerous applications in collaborative and distributed computing like civilian operations (audio/ video conferencing, corporate communications, distance learning, outdoor entertainment activities), emergency search-and-rescue, law enforcement and warfare situations, where establishing and maintaining a communication infrastructure may be expensive or difficult. A common feature among all these applications is one-to-many and many-to-many communications among the participants [2].

Several MANET multicast routing protocols have been proposed in the literature [1]. They are mainly classified as: tree-based and mesh-based protocols. In tree-based protocols, only one route exists between a source and a destination and hence these protocols are efficient in terms of the number of link transmissions. The tree-based protocols can be further divided into two types: source tree-based and shared tree-based. In source tree-based multicast protocols, the tree is rooted at the source. In shared tree-based multicast protocols, the tree is rooted at a core node and all communication between the multicast source and the receiver nodes is through the core node. Even though shared tree-based multicast protocols are more scalable with respect to the number of sources, these protocols suffer under a single point of failure, the core node. On the other hand, source tree-based protocols are more efficient in terms of traffic distribution. In mesh-based multicast protocols, multiple routes exist between a source and each of the receivers of the multicast group. A receiver node receives several copies of the data packets, one copy through each of the multiple paths. Mesh-based protocols provide robustness at the expense of a larger number of link transmissions leading to inefficient bandwidth usage. Considering all the pros and cons of these different classes of multicast routing in MANETs, we feel the source tree-based multicast routing protocols are more efficient in terms of traffic distribution and link usage. Hence, all of our work in this research will be in the category of on-demand source tree-based multicast routing.

Not much work has been done towards the evaluation of MANET multicast routing from a theoretical point of view with respect to metrics such as the hop count per source-receiver path and the number of edges per multicast tree and their impact on the lifetime per multicast tree. These two theoretical metrics significantly contribute and influence the more practically measured performance metrics such as the energy consumption per node, end-to-end delay per data packet, multicast routing overhead and etc. that have been often used to evaluate and compare the different MANET multicast routing protocols in the literature. Hence, we take a different approach in this paper. We study MANET multicast routing using the theoretical algorithms that would yield the benchmarks (i.e., optimum values) for the above two metrics – the Breadth First Search (BFS) algorithm [3] for the minimum hop count per source-receiver path and the minimum Steiner tree heuristic [4] for the minimum number of edges.

Our simulation methodology is outlined as follows: Using the mobility profiles of the nodes gathered offline from a discrete-event simulator (ns-2 [6]), we will generate snapshots of the MANET topology, referred to as Static Graphs, periodically for every fixed time instant. For simulations with a particular algorithm, if a multicast tree is not known for a particular time instant, we will run the algorithm on the static graph in a centralized fashion and adopt the LORA strategy of using this multicast tree as long as it exists for the subsequent static graphs. If the tree no longer exists after a certain time instant, the multicast algorithm is again run to determine a new tree. This procedure is repeated for the entire simulation time. Depending on the algorithm used, the sequence of multicast trees generated either have the minimum hop count per source-receiver path or the minimum number of edges. Our hypothesis is that the multicast trees, determined to optimize one of the two theoretical metrics, would be sub-optimal with respect to the other metric. Through extensive simulation analysis, we confirm our hypothesis to be true and we explain in detail the performance tradeoffs associated with the two metrics.

The rest of the paper is organized as follows: Section 2 reviews the existing related work in the literature. Section 3 introduces the notion of a Static Graph and reviews the BFS algorithm for minimum hop path trees and Kou et al.'s heuristic for minimum edge Steiner trees. Section 4 briefly describes the three mobility models (Random Waypoint, City Section and Manhattan models) simulated in this paper. Section 5 presents the simulation results for the benchmark values of the two theoretical metrics, explores the tradeoffs between these metrics and their impact on the lifetime per multicast tree under each of the three mobility models. Section 6 concludes the paper. For the rest of the paper, the terms 'vertex' and 'node', 'algorithm' and 'heuristic', 'destination' and 'receiver' are used interchangeably. They mean the same.

## 2 Existing Related Work in the Literature

Several MANET multicast routing protocols have been proposed in the literature [1][2]. They are mainly classified as: tree-based and mesh-based protocols. In tree-based protocols, only one route exists between a source and a destination and hence these protocols are efficient in terms of the number of link transmissions. The tree-based protocols can be further divided into two types: source tree-based and shared tree-based. In source tree-based multicast protocols, the tree is rooted at the source. In shared tree-based multicast protocols, the tree is rooted at a core node and all communication between the multicast source and the receiver nodes is through the core node. Even though shared tree-based multicast protocols are more scalable with respect to the number of sources, these protocols suffer under a single point of failure, the core node. On the other hand, source tree-based protocols are more efficient in terms of traffic distribution. In mesh-based multicast protocols, multiple routes exist between a source and each of the receivers of the multicast group. A receiver node receives several copies of the data packets, one copy through each of the multiple paths. Mesh-based protocols provide robustness at the expense of a larger number of link transmissions leading to inefficient bandwidth usage. Considering all the pros and cons of these different classes of multicast routing in MANETs, we feel the source tree-based multicast routing protocols are more efficient in terms of traffic distribution and link usage. Hence, all of our work in this research will be in the category of on-demand source tree-based multicast routing.

Some of the recent performance comparison studies on MANET multicast routing protocols reported in the literature are as follows: In [11], the authors compare the performance of the tree-based MAODV and mesh-based ODMRP protocols with respect to the packet delivery ratio and latency. In [12], the authors propose a stability-based multicast mesh protocol and compare its performance with ODMRP. [13], the authors compare a dominating set-induced mesh based multicast routing protocol for efficient flooding and control overhead and compare the protocol's performance with that of MAODV and ODMRP. In [14], the authors explore the use of genetic algorithms to optimize the performance the performance of tree and mesh based MANET multicast protocols with respect to packet delivery and control overhead. The impact of route selection metrics such as hop count and link lifetime on the performance of on-demand mesh-based multicast ad hoc routing protocols has been examined in [15]. In [16], the author has proposed non-receiver aware and receiver-aware (depending on whether the nodes in the network are aware of the multicast group or not) extensions to the Location Prediction Based Routing (LPBR) protocol to simultaneously minimize the edge count, hop count and number of multicast tree discoveries. An agent-based multicast routing scheme (ABMRS) that uses a set of static and mobile agents for network and multicast initiation and management has been proposed in [17] and compared with MAODV. A zone-based scalable and robust location aware multicast algorithm (SRLAMA) has also been recently proposed for MANETs [18].

## 3 Review of the Theoretical Algorithms used for Multicast Simulations

In this section, we first describe the notion of a static graph, referring to the snapshots of the network topology, on which we run the theoretical algorithms to simulate multicasting. We then describe the two algorithms (BFS and Minimum Steiner tree heuristic) used in this paper.

### 3.1 Static Graph

A static graph is a snapshot of the MANET topology at a particular time instant. Using the mobility profiles of the nodes generated offline from ns-2, we will be able to determine the locations of a node at any particular time instant. A static graph $G(t) = (V, E)$ generated for a particular time instant $t$, comprises of all the nodes in the network as the vertex set $V$; there exists an edge $(u, v) \in E$, if and only if, the Euclidean distance between the two end vertices $u$ and $v \in V$, is less than or equal to the transmission range of the nodes in the network. All the edges in $E$ are of unit weight. We assume a homogeneous network of nodes and all nodes operate at an identical and fixed transmission range.

## 3.2 Breadth First Search (BFS)

The BFS algorithm has been traditionally used to check the connectivity of a network graph. When we start the BFS algorithm on a randomly chosen node, we should be able to visit all the vertices in the graph, if the graph is connected. BFS returns a tree rooted at the chosen start node; when we visit a vertex $v$ for the first time in our BFS algorithm, the vertex $u$ through which we visit $v$ is considered as the predecessor node of $v$ in the tree. Every vertex in the BFS tree, other than the root node, has exactly one predecessor node. When we run BFS on a static graph with unit edge weights, we will be basically obtaining a minimum hop multicast tree such that every node in the graph is connected to the root node (the source node of the multicast group) of the tree on a path with the theoretically minimum hop count.

Figure 1 illustrates BFS in the form of a pseudo code and Figure 2 demonstrates the step-by-step execution of BFS on a sample graph. If $MG \subseteq V$ represents the multicast group – set of receiver nodes and a source node $s$, we start BFS at $s$ and visit all the vertices in the network graph. Once we obtain a BFS tree rooted at $s$, we trace back from every receiver $d \in MG$ and determine the minimum hop $s$-$d$ path. The minimum hop multicast tree is an aggregate of all these minimum hop paths connecting the source $s$ to receiver $d$ in the multicast group.

The set of vertices represented in parentheses below each of the graphs in Figure 2 represents the FIFO-Queue data structure used to maintain the list of vertices that are visited but whose neighbors are yet to be explored (refer the pseudo code in Figure 1). The vertices stored in this queue are extracted in a First-In First-Out fashion and their neighbors are visited if they have not been already explored. Note that, for simplicity, we restrict our research in this paper to only single source multicast groups; the research could be easily extended for multicast groups with more than one source node. Once we establish the benchmarks for single source multicast groups in this paper, we will extend the research for multi-source multicast groups in the immediate future.

---

**Input:** Static Graph $G = (V, E)$, source $s$
**Auxiliary Variables/Initialization:** Nodes-Explored = $\Phi$, FIFO-Queue = $\Phi$, root-node
$\forall v \in V$, Predecessor(v) = NULL
**Begin** Algorithm *BFS (G, s)*
  root-node = $s$
  Nodes-Explored = Nodes-Explored U {root-node}
  FIFO-Queue = FIFO-Queue U {root-node}
  **while** ( |FIFO-Queue| > 0 ) **do**
    first-node $u$ = Dequeue(FIFO-Queue) // extract the first node
    **for** (every edge $(u, v) \in E$) **do** // i.e. every neighbor v of node u
      **if** ($v \notin$ Nodes-Explored) **then**
        Nodes-Explored = Nodes-Explored U {$v$}
        FIFO-Queue = FIFO-Queue U {$v$}
        Predecessor ($v$) = $u$
      **end if**
    **end for**
  **end while**

**End** Algorithm *BFS*

---

**Figure 1:** Pseudo Code for Breadth First Search (BFS)

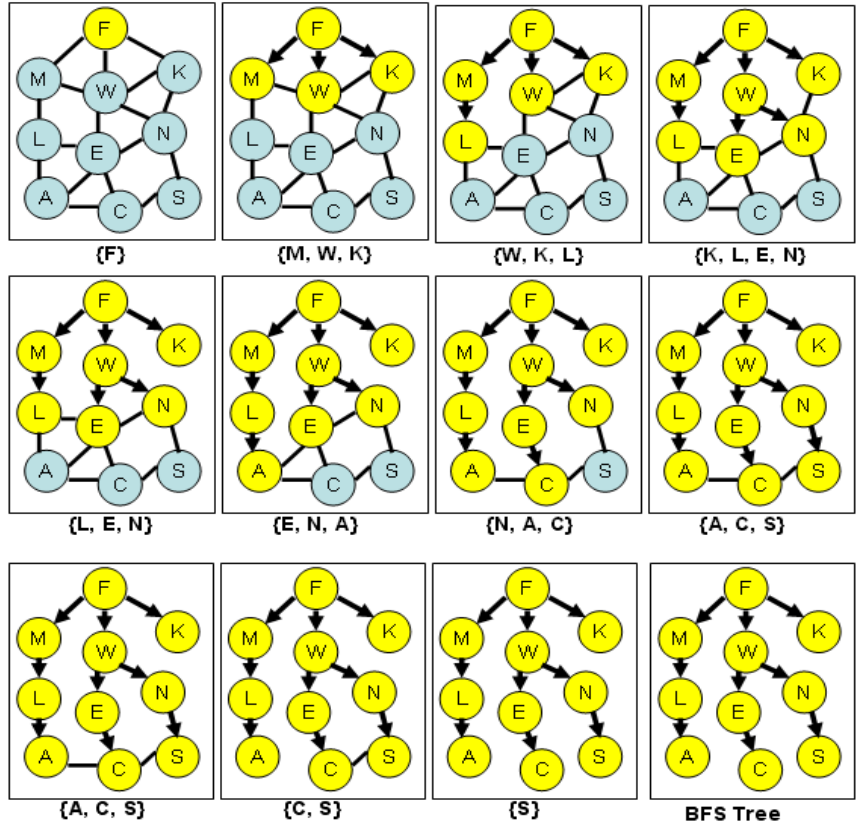

**Figure 2:** Execution of BFS on a Sample Graph

### 3.3 Minimum Edge Multicast Steiner Tree

Given a static graph, $G = (V, E)$, where $V$ is the set of vertices, $E$ is the set of edges and a subset of vertices (called the multicast group or Steiner points) $MG \subseteq V$, the multicast Steiner tree is the tree with the least number of edges required to connect all the vertices in $MG$. Unfortunately, the problem of determining a minimum edge Steiner tree in an undirected graph like that of the static graph is NP-complete. Efficient heuristics (e.g., [4]) have been proposed in the literature to approximate a minimum Steiner tree. In this paper, we use the Kou et al's [4] well-known $O(|V||MG|^2)$ heuristic ($|V|$ is the number of nodes in the network graph and $|MG|$ is the size of the multicast group comprising of the source nodes and the receiver nodes) to approximate the minimum edge Steiner tree in graphs representing snapshots of the network topology. An *MG-Steiner-tree* is referred to as the minimum edge Steiner tree connecting the set of nodes in the multicast group $MG \subseteq V$. In unit disk graphs such as the static graphs used in our research, Step 5 of the heuristic is not needed and the minimal spanning tree $T_{MG}$ obtained at the end of Step 4 could be considered as the minimum edge Steiner tree.

---

**Input:** A Static Graph $G = (V, E)$
Multicast Group $MG \subseteq V$
**Output:** A *MG-Steiner-tree* for the set $MG \subseteq V$

**Begin** Kou et al Heuristic ($G$, $MG$)
  **Step 1:** Construct a complete undirected weighted graph $G_C = (MG, E_C)$ from $G$ and $MG$ where $\forall (v_i, v_j) \in E_C$, $v_i$ and $v_j$ are in $MG$, and the weight of edge $(v_i, v_j)$ is the length of the shortest path from $v_i$ to $v_j$ in $G$.

**Step 2:** Find the minimum weight spanning tree $T_C$ in $G_C$ (If more than one minimal spanning tree exists, pick an arbitrary one).
**Step 3:** Construct the sub graph $G_{MG}$ of $G$, by replacing each edge in $T_C$ with the corresponding shortest path from $G$ (If there is more than one shortest path between two given vertices, pick an arbitrary one).
**Step 4:** Find the minimal spanning tree $T_{MG}$ in $G_{MG}$ (If more than one minimal spanning tree exists, pick an arbitrary one). Note that each edge in $G_{MG}$ has weight 1.

**return** $T_{MG}$ as the *MG-Steiner-tree*

**End** Kou et al Heuristic

**Figure 3:** Kou et al's Heuristic [4] to find an Approximate Minimum Edge Steiner Tree

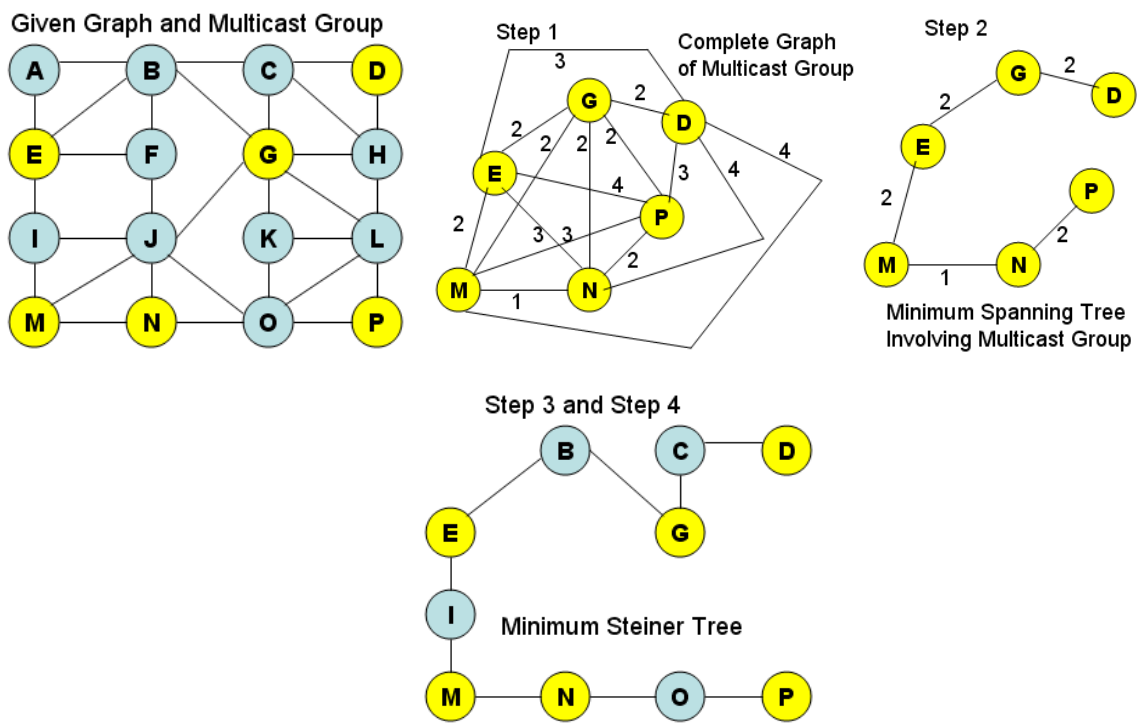

**Figure 4:** Example to Illustrate the Construction of a Minimum Steiner Tree

We give a brief outline of the heuristic in Figure 3 and illustrate the working of the heuristic through an example in Figure 4. The vertices {D, G, E, M, N, P} form the multicast group in the vertex set {A, B … P}. As observed in the example, the subgraph $G_{MG}$ obtained in Step 3 is nothing but the minimal spanning tree $T_{MG}$, which is the output of Step 4. In general, for unit disk graphs, like the static graphs we are working with, the outputs of both Steps 3 and 4 are the same and it is enough that we stop at Step 3 and output the MG-Steiner-tree.

## 4  Review of the Mobility Models

In this section, we provide a brief overview of the Random Waypoint mobility model [8] commonly used in MANET simulation studies and the widely used mobility models for vehicular ad hoc networks (VANETs), viz., City Section [9] and Manhattan mobility models [10]. All the three mobility models assume the network to be confined within fixed boundary conditions. The mobility of a node is independent of the other nodes in all the three mobility models. Under the Random Waypoint model, each node can move anywhere within a network region. For the two VANET models, the network is assumed to be divided into grids of square blocks with identical block length. The network for the City Section and Manhattan models is thus basically composed of a number of horizontal and vertical streets with each street having two lanes, one for each direction (east and west direction for horizontal streets; north and south direction for vertical streets); nodes can move only along the grids of horizontal and vertical streets. All streets are assumed to have identical value for the maximum speed limit ($v_{max}$).

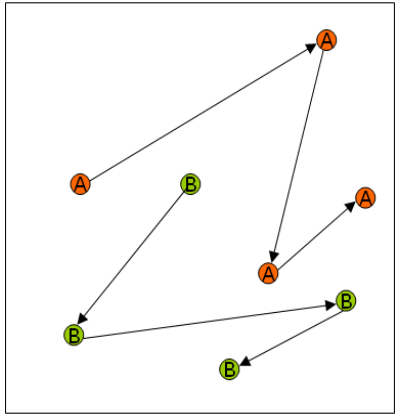 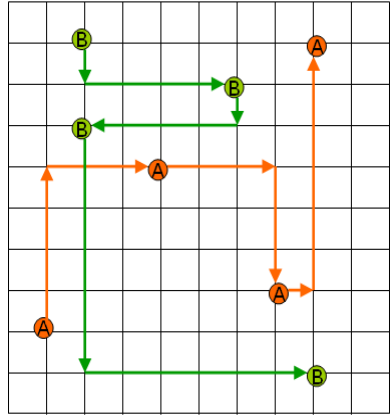 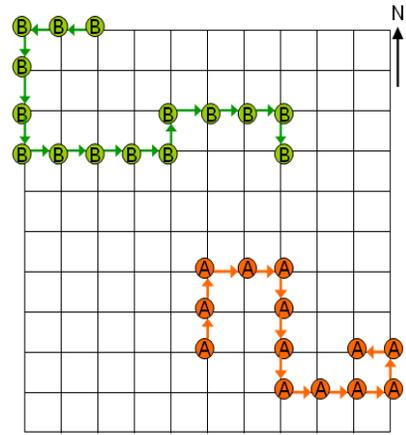

**Figure 5.** Movement under the Random Waypoint Mobility Model

**Figure 6.** Movement under the City Section Mobility Model

**Figure 7.** Movement under the Manhattan Mobility Model

### 4.1 Random Waypoint Mobility Model

The nodes are initially assumed to be placed at random locations in the network. As each node moves independently of the other nodes in the network, the mobility pattern described here is applicable for every node. The movement of a node is described as follows: The node randomly chooses a target location (within the network) to move. The velocity to move to the chosen location is uniform randomly chosen from the interval [$v_{min}$,…,$v_{max}$]. The node is assumed to move on a straight line to the chosen location with the chosen velocity. After reaching the targeted location, the node may stay there for a certain time, called the pause time, and then continues to move by choosing a different target location (that is independent of the current and previous locations) under a different randomly chosen velocity from the interval [$v_{min}$,…,$v_{max}$]. Each time a node changes direction, it is referred to as moving to a new waypoint. Figure 5 illustrates the mobility of two nodes (*A* and *B*) moving in random directions with randomly chosen velocities anywhere within a network. In our simulations in this paper, the values of both $v_{min}$ and the pause time are 0.

### 4.2 City Section Mobility Model

To start with, each node is placed at a randomly chosen street intersection. Note that it is possible for more than one node to be placed at a particular street intersection. The mobility of a node is described as follows: Each node chooses a random street intersection (within the grid network) to move with a velocity uniform-randomly chosen from the range [0, …, $v_{max}$]. The node then moves to the chosen street intersection with the chosen velocity on a path that will incur the least amount of travel time. Any tie, between two or more paths that offer the same minimum travel time, is broken arbitrarily. After moving to the chosen street intersection, the node may stay there for a pause time (in our simulations, there is zero pause time) and then continues to move by randomly choosing a different target street intersection

(independent of the current and previous street intersections) under a different uniform-randomly chosen velocity form the range [0, …, $v_{max}$]. The above procedure is independently repeated by each node. Figure 6 illustrates the movement of two nodes (*A* and *B*) under the City Section model.

### 4.3 Manhattan Mobility Model

The nodes are initially assumed to be placed in randomly chosen street intersections. The mobility of a node is decided one street block at a time. In Figure 7, to start with, node A has equal chance (25% each) to move in each of the four possible directions (east, west, north or south) starting from its initial location; whereas, node B can move only either to the west, east or south with a 1/3 chance for each direction. The velocity at which a node moves from one street to the subsequent street intersection is uniform-randomly chosen from the range [0, …, $v_{max}$]. After a node moves to the chosen neighboring street intersection, the subsequent street intersection to which the node will move is chosen probabilistically. If a node can continue to move in the same direction or can also change directions, the node has 50% chance of moving in the same direction; 25% chance to turn on either side, with the exact new direction depending on the direction of the previous movement. If a node has only two options, then the node either moves to the next street intersection by continuing in the same direction or changes direction. For example, in Figure 7, after node *A* reaches the rightmost network boundary, it can either move to the north or to the south, each with a probability of 0.5 and the node chooses to move in the north direction. After moving to the next street intersection, node *A* can continue to move northwards or turn left and move eastwards, each with a probability of 0.5. If a node has only one option to move (this situation occurs when a node reaches any of the four corners forming the network boundary), then the node has no other choice except to explore that option. For example, in Figure 7, the only option for node B, which was initially traveling westward and reaching the corner of the network, is to turn to the left and proceed southwards.

### 5   Simulations

The simulations have been conducted in a discrete-event simulator implemented by the author in Java. The two multicast algorithms have been implemented in a centralized fashion. We generate the static graphs by taking snapshots of the network topology, periodically for every 0.25 seconds, and run the two multicast algorithms. The simulation time is 1000 seconds. We consider a square network of dimensions 1000m x 1000m. The transmission range of the nodes is 250m. The network density is varied by performing the simulations with 50 nodes (low density) and 100 nodes (high density). We assume there is only one source for the multicast group and three different values for the number of receivers per multicast group are considered: 3 (small), 10 (moderate) and 18 (large). A multicast group comprises of a source node and a list of receiver nodes, the size of which is mentioned above. The $v_{max}$ values used for each of the three mobility models (Random Waypoint, City Section and Manhattan models) are 5 m/s (low mobility), 25 m/s (moderate mobility) and 50 m/s (high mobility). The pause time is 0 seconds. The reader is referred to Section 4 for a detailed description on the behavior of the mobility models.

   The performance metrics measured are as follows. Each performance metric illustrated in Figures 8 through 17 is measured using 5 different lists of receiver nodes for the same size and the multicast algorithm is run on five different mobility trace files generated for a particular value of $v_{max}$ for each mobility model:
(i) Tree Connectivity: This metric refers to the percentage of time instants there exists a multicast tree connecting the source node to the receiver nodes of the multicast group, averaged over the mobility profiles generated for a particular value of $v_{max}$ for a given number of network nodes and number of receivers per multicast group.
(ii) Lifetime per Multicast Tree: Whenever a link break occurs in a multicast tree, we establish a new multicast tree. The lifetime per multicast tree is the average of the time between successive multicast tree discoveries for a particular routing protocol or algorithm, over the duration of the multicast session. The larger the value of the lifetime per multicast tree, the lower the number of multicast tree transitions or discoveries needed.
(iii) Number of links per tree: This metric refers to the total number of links in the entire multicast tree, time-averaged over the duration of the multicast session. For example, a multicast session uses two trees, one tree with 10 links for 3 seconds and another tree with 15 links for 6 seconds, then the time-averaged value for the number of links per tree for the 9-second duration of the multicast session is (10*3 + 15*6)/(3 + 6) = 13.3 and not 12.5.
(iv) Number of hops per receiver: We measure the number of hops in the paths from the source to each receiver of the multicast group and average it for the duration of the multicast session. This metric is also a time-averaged value of the number of hops from a multicast source to a receiver and then averaged over all the receivers of a multicast session.

### 5.1 Tree Connectivity

The connectivity of the trees (refer Figure 8) does not depend on any individual multicast algorithm used and depends only on the mobility model, network density, node mobility and the number of receivers per multicast group. The Manhattan model incurs the lowest tree connectivity for most of the scenarios, especially for those with low network density (number of nodes) and larger multicast group size. On the other hand, the Random Waypoint model incurs the largest tree connectivity.

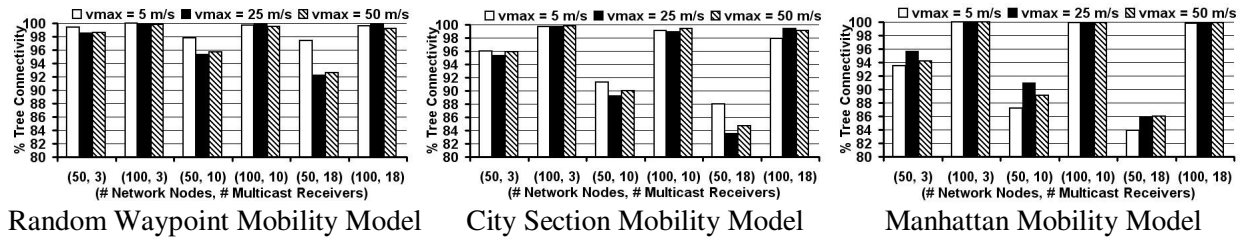

Random Waypoint Mobility Model      City Section Mobility Model      Manhattan Mobility Model

**Figure 8:** Percentage Tree Connectivity under the Different Mobility Models

For a fixed density and node mobility, as we increase the number of receivers per multicast group, the number of time instants for which we could connect the source node to all the receiver nodes decreases. With node mobility, the source may not be connected all the time to all the receivers. The probability of the source connected to all the receiver nodes decreases with increase in the number of receivers per multicast group. On the other hand, for a fixed node mobility and number of receivers per multicast group, the connectivity of a multicast tree increases with increase in the network density. This could be attributed to the availability of a larger number of nodes to connect the source node to the multicast receivers. For low density networks, we observe that as the number of receivers per multicast group increases, the percentage of tree connectivity decreases with increase in maximum node velocity. This can be attributed to an appreciable probability (in low density networks) of not being able to find a path that connects a source node to all the receiver nodes of the multicast group. As the network density increases, we do not observe relatively less variations in tree connectivity with respect to increase in the number of receivers per multicast group as well as with increase in maximum node velocity.

### 5.2 Number of Edges per Multicast Tree and Hop Count per Source-Receiver Path

As expected, the minimum-edge based Steiner trees incurred the smallest number of edges per multicast trees. In most of the scenarios, the number of edges per multicast tree under a Random Waypoint model is larger than that incurred with the City Section model, which is larger than that incurred with the Manhattan model. On average, the number of edges per minimum hop tree is 13-35% more than those incurred with the minimum edge tree. With an objective to optimize the hop count, minimum hop based multicast trees select edges that could constitute a minimum hop path, but with a higher probability of failure in the immediate future. The physical distance between the constituent nodes of an edge on a minimum hop path is close to the transmission range of the nodes at the time of tree formation itself. For a given network density, as we increase the number of receivers per multicast group from 3 to 18, the average number of edges per multicast tree increased by a factor of 3 to 4. For the minimum hop and minimum edge trees, for a given level of node mobility and number of receivers per group, as we increase the network density, the number of edges per tree remains the same or only slightly decreases.

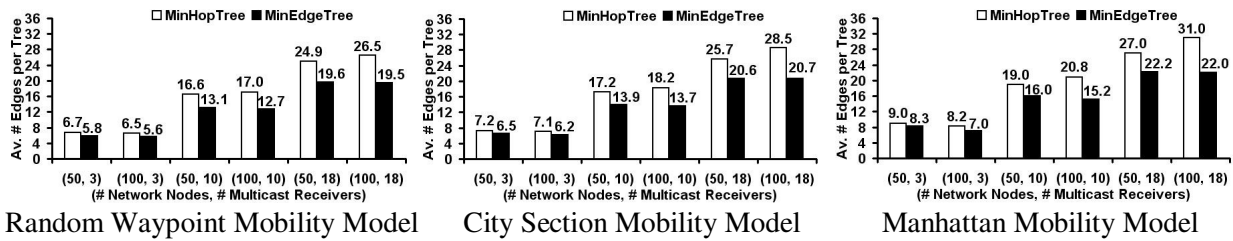

Random Waypoint Mobility Model    City Section Mobility Model    Manhattan Mobility Model

**Figure 9:** Average # Edges per Tree under the Different Mobility Models (Max. Node Velocity: 5 m/s)

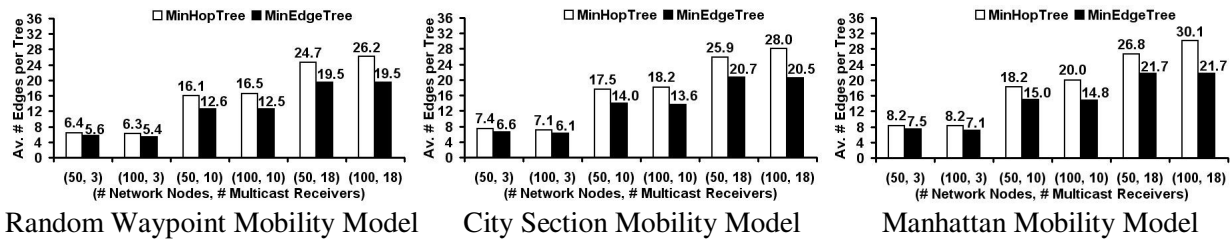

Random Waypoint Mobility Model    City Section Mobility Model    Manhattan Mobility Model

**Figure 10:** Average # Edges per Tree under the Different Mobility Models (Max. Node Velocity: 25 m/s)

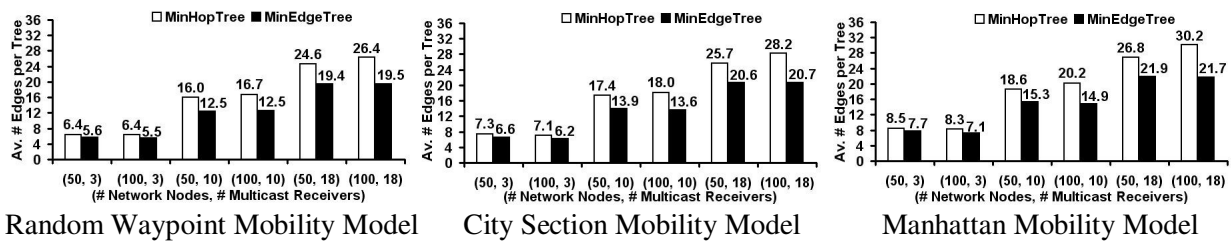

Random Waypoint Mobility Model    City Section Mobility Model    Manhattan Mobility Model

**Figure 11:** Average # Edges per Tree under the Different Mobility Models (Max. Node Velocity: 50 m/s)

As expected, the minimum hop multicast trees incurred the lowest hop count per source-receiver path. In most of the scenarios, the hop count per source-receiver path for both the multicast trees incurred under a Random Waypoint model is larger than that incurred with the City Section model, which is larger than that incurred with the Manhattan model. The larger hop count per source-receiver path for minimum edge trees could be attributed to a relatively lower number of edges compared to the minimum hop trees. As we connect the source node to the multicast receivers with the lowest possible number of edges, the number of hops between the source node and to each of the receiver nodes increases. This is the tradeoff between the objectives of minimizing the number of edges per multicast tree and the hop count per individual source-receiver paths in the multicast tree.

For both minimum hop and minimum edge multicast trees, for a given network density and number of receivers per multicast group, there is appreciably no impact of the maximum node velocity on the average number of edges per tree as well as the hop count per source-receiver path. For a given level of node mobility (i.e., maximum node velocity) and network density, as we increase the number of receivers per multicast group, the average hop count per source-receiver path for minimum hop trees decreases. On the other hand, the average hop count per source-receiver path for minimum edge trees increases. This could be attributed to the relatively fewer number of edges in the minimum edge trees compared to those incurred by the minimum hop trees. The relatively more edges in minimum hop trees at larger number of receivers per multicast group results in lower hops count per source-receiver path. The average number of edges per minimum hop tree for a network of 50 nodes and 3 receivers per multicast group is about 1 edge more than those incurred by the minimum edge trees; on the other hand, the average number of edges per minimum hop tree for a network of 50 nodes and 18 receivers per multicast group is about 7 edges more than the minimum. Similar observations could be made for network of 100 nodes.

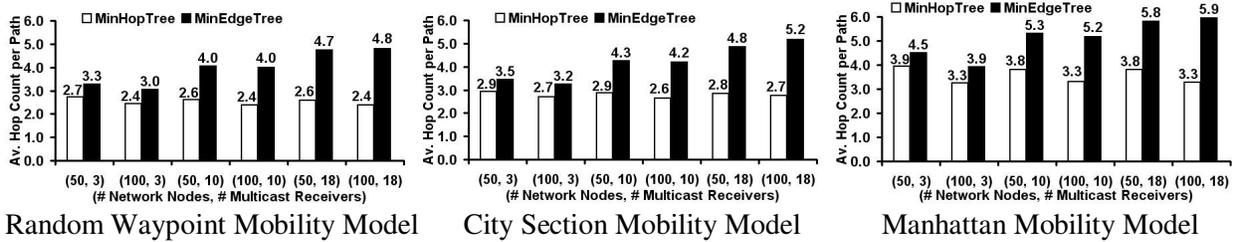

Random Waypoint Mobility Model | City Section Mobility Model | Manhattan Mobility Model

**Figure 12:** Average Hop Count / Path under the Different Mobility Models (Max. Node Velocity: 5 m/s)

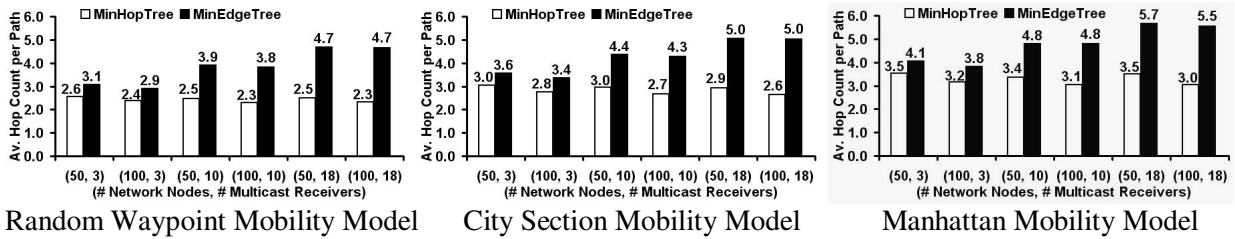

Random Waypoint Mobility Model | City Section Mobility Model | Manhattan Mobility Model

**Figure 13:** Average Hop Count / Path under the Different Mobility Models (Max. Node Velocity: 25 m/s)

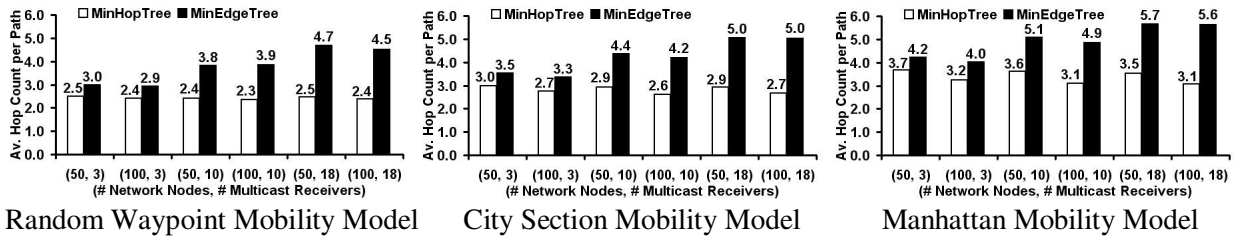

Random Waypoint Mobility Model | City Section Mobility Model | Manhattan Mobility Model

**Figure 14:** Average Hop Count / Path under the Different Mobility Models (Max. Node Velocity: 50 m/s)

When compared to the average hop count per source-receiver path incurred by minimum hop trees, the average hop count per source-receiver path for minimum edge trees is 20% (for smaller number of receivers per multicast group) to 100% (for larger number of receivers per multicast group) more. Note that with increase in the network density and/or the number of receivers per multicast group, the trend of the hop counts per source-receiver path for minimum hop trees is to decrease; whereas, the trend of the hop count per source-receiver path for minimum edge trees is to increase. The hop count per source-receiver path for minimum hop trees decreases by at most 14% and 30% respectively; whereas, the hop count per source-receiver path for minimum edge trees increases by at most 47%.

## 5.3 Lifetime per Multicast Tree

The minimum edge multicast trees had a relatively longer lifetime compared to the minimum hop multicast trees. This could be attributed to (i) the increased number of edges (refer to Section 5.2 for more on this observation) in a minimum hop multicast tree; (ii) the physical Euclidean distance between the constituent nodes of an edge on a minimum hop path is close to the transmission range of the nodes at the time of tree formation itself. Thus, the probability of an edge failure is quite high at the time of formation of the tree; (iii) the edges of a tree are also independent from each other. All these three factors play a significant role in the relatively lower lifetime per minimum hop multicast tree. While both the minimum hop and minimum edge trees exist for a relatively longer time under the Manhattan mobility model; the lifetime of the trees was the least under the City Section model for most of the scenarios.

For both the multicast algorithms, for a fixed network density, as the number of receivers per multicast group is increased, the lifetime per multicast tree decreases moderately at low node mobility and decreases drastically (as large as one-half to one-third of the value at smaller number of receivers per group) at moderate and high node mobility scenarios. This could be attributed to the difficulty in finding a tree that would keep the source node connected to the receivers of the multicast group for a longer time, with increase in node mobility and/or the number of receivers per multicast group. For a given number of receivers per multicast group and node mobility, the lifetime per minimum hop trees and minimum edge trees slightly decreases as we double the network density. The decrease is more predominant for minimum hop trees and this could be attributed to the relatively unstable minimum hop paths in high density networks (refer Section 3.2 for more discussion on this observation).

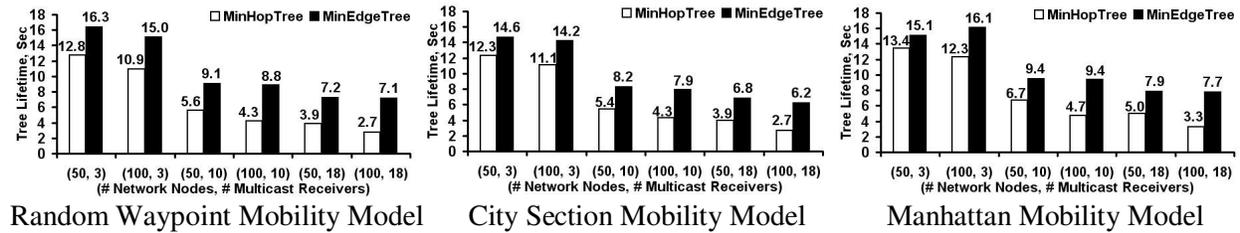

**Figure 15:** Average Tree Lifetime under the Different Mobility Models (Max. Node Velocity: 5 m/s)

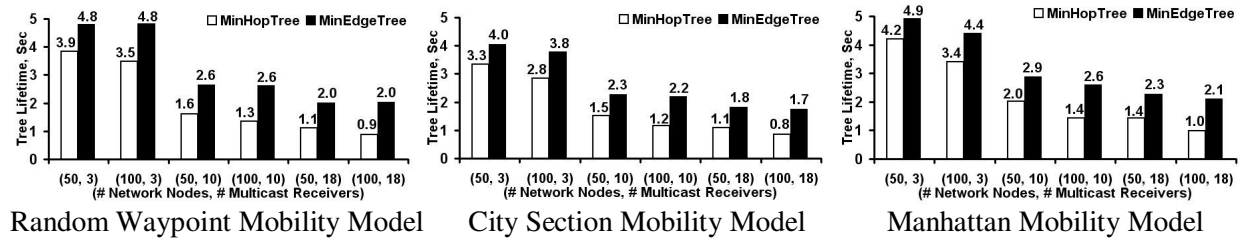

**Figure 16:** Average Tree Lifetime under the Different Mobility Models (Max. Node Velocity: 25 m/s)

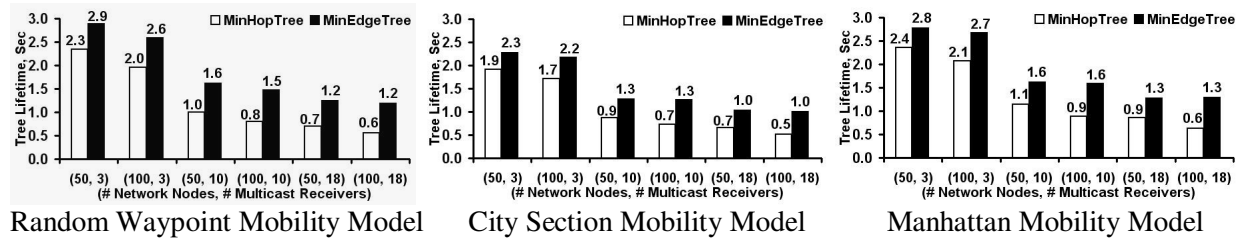

**Figure 17:** Average Tree Lifetime under the Different Mobility Models (Max. Node Velocity: 50 m/s)

For a given level of node mobility, the lifetime per minimum edge tree is 23% (low density) to 38% (high density); 61% (low density) to 107% (high density) and 76% (low density) to 160% (high density) larger than the lifetime per minimum hop tree for small, moderate and larger number of receivers per multicast group respectively. For both minimum hop and minimum edge trees, for a given network density and number of receivers per group, as we increase the maximum node velocity to 25 m/s and 50 m/s, the lifetime per tree reduces by 1/3$^{rd}$ to 1/6$^{th}$ of their value at a maximum node velocity of 5 m/s.

## 6  Conclusions

We have described the algorithms that can be used to obtain benchmarks for the minimum hop count per source-receiver path and minimum number of edges per tree for multicast routing in mobile ad hoc networks. Simulations have been conducted to obtain such benchmarks for different conditions of network density, node mobility and number of receivers per multicast group under three different mobility models – the Random Waypoint model (used for MANETs) plus the City Section and Manhattan model (used for VANETs). Both the minimum edge and minimum hop based multicast trees are inherently more stable under the Manhattan model and least stable under the City Section model. The Random Waypoint model supports the minimum edge trees and minimum hop trees to have the lowest values for the number of edges and hop count per source-receiver path metrics.

For a particular mobility model, the minimum hop based multicast trees have a larger number of edges than the theoretical minimum – the minimum hop trees are unstable and their lifetime decreases with increase in the number of edges. This could be attributed to the instantaneous decision taken by the minimum hop path algorithm to select a tree without any consideration for the number of edges and their lifetime. The minimum edge trees have a relatively larger hop count per source-receiver path and the hop count per path increases with the number of receivers per multicast group. The relatively fewer edges in the minimum edge tree results in a relatively larger lifetime compared to the minimum hop trees, as each edge in these two trees are independent. The simulation results thus indicate a complex tradeoff between the hop count per source-receiver paths and number of edges per tree vis-à-vis their impact on the lifetime per tree for multicast routing.

## References

[1] C. Siva Ram Murthy and B. S. Manoj, "Routing Protocols for Ad Hoc Wireless Networks," *Ad Hoc Wireless Networks: Architectures and Protocols*, Chapter 7, pp. 299 – 364, Prentice Hall, 1$^{st}$ Edition, June 2004.
[2] C. K. Toh, G. Guichal and S. Bunchua, "ABAM: On-demand Associatvity-based Multicast Routing for Ad hoc Mobile Networks," *Proceedings of the 52$^{nd}$ IEEE VTS Fall Vehicular Technology Conference*, Vol. 3, pp. 987 – 993, September 2000.
[3] T. H. Cormen, C. E. Leiserson, R. L. Rivest and C. Stein, "Introduction to Algorithms," 2$^{nd}$ Edition, MIT Press/ McGraw-Hill, Sept. 2001.
[4] L. Kou, G. Markowsky and L. Berman, "A Fast Algorithm for Steiner Trees," Vol 15, pp. 141-145, *Acta Informatica*, Springer-Verlag, 1981.
[5] N. Meghanathan, "On the Stability of Paths, Steiner Trees and Connected Dominating Sets in Mobile Ad hoc Networks," *Ad Hoc Networks*, Vol. 6, No. 5, pp. 744-769, July 2008.
[6] K. Fall, K. Varadhan, "The ns Manual," The VINT Project, A Collaboration between researchers at UC Berkeley, LBL, USC/ISI and Xerox PARC.
[7] A. Farago and V. R. Syrotiuk, "MERIT: A Scalable Approach for Protocol Assessment," *Mobile Networks and Applications*, Vol. 8, No. 5, pp. 567 – 577, October 2003.
[8] C. Bettstetter, H. Hartenstein and X. Perez-Costa, "Stochastic Properties of the Random-Way Point Mobility Model," *Wireless Networks*, pp. 555 – 567, Vol. 10, No. 5, September 2004.
[9] T. Camp, J. Boleng and V. Davies, "A Survey of Mobility Models for Ad Hoc Network Research," Wireless Communication and Mobile Computing, Vol. 2, No. 5, pp. 483-502, September 2002.
[10] F. Bai, N. Sadagopan and A. Helmy, "IMPORTANT: A Framework to Systematically Analyze the Impact of Mobility on Performance of Routing Protocols for Ad hoc Networks," *Proceedings of the IEEE International Conference on Computer Communications*, pp. 825-835, March-April, 2003.

[11] A. Vasiliou and A. A. Economides, "Evaluation of Multicast Algorithms in MANETs," *Proceedings of the 3rd World Enformatika Conference – International Conference on Telecommunications and Electronic Commerce*, vol. 5, pp. 94-97, Stevens Point, Wisconsin, USA, 2005.
[12] R. Biradar, S. Manvi and M. Reddy, "Mesh Based Multicast Routing in MANETs: Stable Link Based Approach," *International Journal of Computer and Electrical Engineering*, vol. 2, no. 2, pp. 371-380, April 2010.
[13] E. Menchaca-Mendez, R. Vaishampayan, J. J. Garcia-Luna-Aceves and K. Obraczka, "DPUMA: A Highly Efficient Multicast Routing Protocol for Mobile Ad Hoc Networks," *Lecture Notes in Computer Science*, vol. 3738, pp. 178-191, 2005.
[14] E. Baburaj and V. Vasudevan, "Exploring Optimized Route Selection Strategy in Tree- and Mesh-Based Multicast Routing in MANETs," *International Journal of Computer Applications in Technology*, vol. 35, no. 2-4, pp. 174-182, 2009.
[15] N. Meghanathan and S. Vavilala, "Impact of Route Selection Metrics on the Performance of On-Demand Mesh-based Multicast Ad hoc Routing," *Computer and Information Science*, vol. 3, no. 2, pp. 3-18, May 2010.
[16] N. Meghanathan, "Multicast Extensions to the Location-Prediction Based Routing Protocol for Mobile Ad hoc Networks," *Lecture Notes of Computer Science*, vol. 5682, pp. 190-199, August 2009.
[17] S. S. Manvi and M. S. Kakkasageri, "Multicast Routing in Mobile Ad hoc Networks by using a Multiagent System," *Elsevier Information Sciences*, vol. 178, no. 6, pp. 1611-1628, March 2008.
[18] P. Kamboj and A. K. Sharma, "Scalable and Robust Location Aware Multicast Algorithm (SRLAMA) for MANET," *International Journal of Distributed and Parallel Systems*, vol. 1, no. 2, pp. 10-24, November 2010.